\documentclass[preprint,journal]{vgtc}            

\DeclareGraphicsExtensions{.png,PNG,.pdf,.PDF}

\onlineid{1645}

\vgtccategory{Research}

\vgtcpapertype{Applications}

\graphicspath{{figs/}{figures/}{pictures/}{images/}{./}}
\usepackage{multirow}
\usepackage{color}
\usepackage{lipsum}                    
\usepackage{amssymb,amsmath}
\usepackage{bbm} 
\usepackage{tabularx} 
\usepackage{wrapfig}
\usepackage{booktabs}
\usepackage{makecell}
\usepackage{graphicx}
\usepackage{dblfloatfix}

\usepackage{mathptmx}

\usepackage{tabu}                      %
\usepackage{booktabs}                  %
\usepackage{lipsum}                    %
\usepackage{mwe}                       %
\usepackage{tabularx}
\usepackage{ragged2e}
\usepackage{array}
\usepackage{multirow}
\usepackage{amsmath} 
\usepackage{bm}
\usepackage{threeparttable}
\usepackage{graphicx}

\usepackage{mathptmx}   

\title{TrajLens: Visual Analysis for Constructing Cell Developmental Trajectories in Cross-Sample Exploration}
\author{%
  \authororcid{Qipeng Wang}{0009-0001-0350-4701},
\authororcid{Shaolun Ruan}{0000-0002-6163-9786},
\authororcid{Rui Sheng}{0000-0001-9321-6756},
\authororcid{Yong Wang}{0000-0002-0092-0793}, 
Min Zhu,
and \authororcid{Huamin Qu}{0000-0002-3344-9694}
}

\authorfooter{
    \item Q. Wang, M. Zhu are with Sichuan University. E-mail: wangqipengscu@stu.scu.edu.cn, zhumin@scu.edu.cn
  \item S. Ruan is with Singapore Management University. E-mail: slruan.2021@phdcs.smu.edu.sg. S. Ruan is also with Monash University.
  \item R. Sheng, H. Qu are with Hong Kong University of Science and Technology. E-mail: rshengac@connect.ust.hk, huamin@cse.ust.hk.
  \item Y. Wang is with Nanyang Technological University. E-mail: yong-wang@ntu.edu.sg.
  \item M. Zhu is the corresponding author.
}

\abstract{
	Constructing cell developmental trajectories is a critical task in single-cell RNA sequencing (\scrna) analysis, enabling the inference of potential cellular progression paths.
	However, current automated methods are limited to establishing cell developmental trajectories within individual samples, necessitating biologists to manually link cells across samples to construct complete cross-sample evolutionary trajectories that consider cellular spatial dynamics.
	This process demands substantial human effort due to the complex spatial correspondence between each pair of samples.
	To address this challenge, we first proposed a GNN-based model to predict cross-sample cell developmental trajectories. 
	We then developed \system, a visual analytics system that supports biologists in exploring and refining the cell developmental trajectories based on predicted links.
	Specifically, we designed the visualization that integrates features on cell distribution and developmental direction across multiple samples, providing an overview of the spatial evolutionary patterns of cell populations along trajectories. Additionally, we included contour maps superimposed on the original cell distribution data, enabling biologists to explore them intuitively.
	To demonstrate our system's performance, we conducted quantitative evaluations of our model with two case studies and expert interviews to validate its usefulness and effectiveness.
	}


\teaser{
  \centering
  \includegraphics[width=1.0\linewidth]{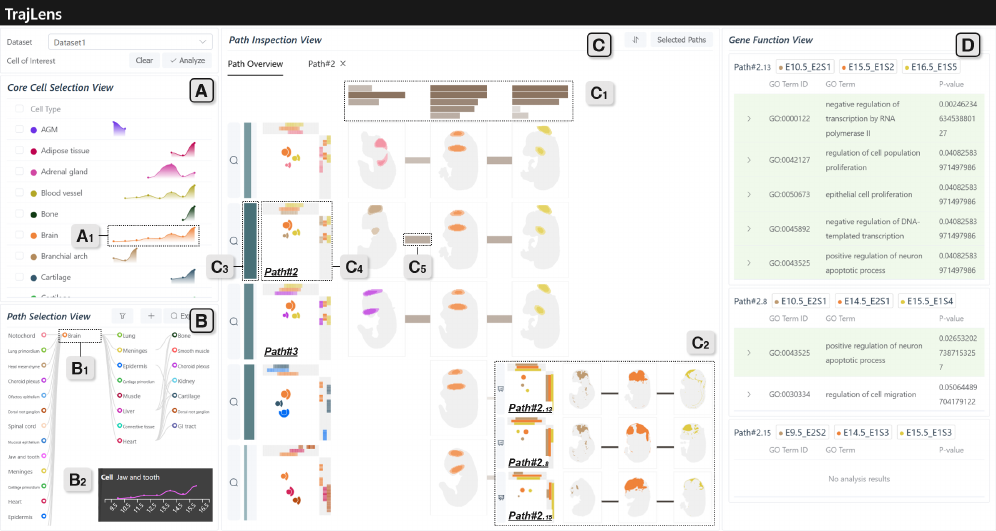}
  \caption{\system: 
  	(A) The Core Cell Selection View helps users select the core cell type based on temporal occurrences and quantitative changes during biological development.
  	(B) The Path Selection View assist users in intuitively selecting and analyzing complex cell developmental paths with a focus on high-frequency and diverse trajectories.
  	(C) The Path Inspection View reveals spatial evolutionary patterns through a multi-row synchronized view, facilitating comprehensive validation of biological evolutionary relationships with features for sequence inspection, single-sample inspection, and similarity assessment.
  	(D) The Gene Function View present significant gene functions identified through the analysis of single-sample sequences.
  }
  \label{fig1}
}

\newcommand{\ie}{i.e.{\xspace}}
\newcommand{\eg}{e.g.,\xspace}

\newcommand{\etal}{{\xspace}et~al.{\xspace}}
\newcommand{\etc}{etc.{\xspace}}

\newcommand{\Ex}[1]{E{\small{#1}}\xspace}

\newcommand{\Path}[1]{\texttt{Path\#{#1}}{\xspace}}
\newcommand{\Subpath}[2]{\texttt{Path\#{#1}.\small{#2}}}

\newcommand{\Fig}[2]{(Fig.{#1}(#2))}
\newcommand{\Subfig}[3]{(Fig.{#1}(#2{\small{#3}}))}

\newcommand{\scrna}{scRNA-seq\xspace}

\newcommand{\system}{\textit{TrajLens}\xspace}
\newcommand{\celltable}{\textit{Core Cell Selection View}\xspace}
\newcommand{\pathselection}{\textit{Path Selection View}\xspace}
\newcommand{\pathcomparison}{\textit{Path Inspection View}\xspace}
\newcommand{\pathprojection}{\textit{Path Summary View}\xspace}
\newcommand{\trajectorycomparison}{\textit{Trajectory Inspection View}\xspace}

\newcommand{\genefunction}{\textit{Gene Function View}\xspace}

\newcolumntype{Y}{>{\RaggedRight\arraybackslash}X} %
\newcolumntype{Z}{>{\Centering\arraybackslash}X}

\begin{document}

\firstsection{Introduction}

\maketitle

In the field of single-cell RNA sequencing (\scrna), constructing cross-sample cell developmental trajectories is a crucial analytical task that helps reveal the progression of cells through multiple developmental stages \cite{dtflow2021wei}. 
These trajectories provide insights into cellular differentiation, thereby advancing our understanding of biological development \cite{developmental2024singh}, disease progression \cite{mechanism2021kang}, and targeted therapy \cite{therapy2024yang}.
The construction process typically involves two key phases: first, establishing cellular trajectories merely within single biological sample and then linking these trajectories across multiple samples to reveal cross-sample trajectories.
While current computational tools like Monocle \cite{monocle2019cao} and PAGA \cite{paga2019wolf} have automated the first phase by effectively modelling intra-sample cellular dynamics, they lack capabilities for cross-sample connection.
Consequently, biologists often make every effort to manually connecting these trajectories between samples, which is subjective and prone to errors.

Specifically, constructing and exploring cell developmental trajectories across multiple samples presents significant challenges. 
Currently, biologists must manually analyze which cell populations share the same expressed genes and connect them, thereby creating a many-to-many mapping problem that complicates the analytic task.
Additionally, experts need to assess the spatial distribution correspondence of cell populations in these trajectories, which is complicated by their variable distributions across samples. 
Consequently, it is difficult to identify developmental continuities and biological evolutionary relationships of these trajectories accurately, which highlights the need for automated tools to improve the reliability of cross-sample trajectory connections.

However, developing such tools requires resolving two critical aspects. 
First, developing a computational model that integrates multi-dimensional cellular features, such as gene expression, spatial correspondence, and cross-sample dynamics, is essential for predicting developmental trajectories between cell populations across samples. 
Moreover, it is particularly challenging for a model to capture cross-sample dynamics due to its computational complexity.
Second, biologists need to manually understand and refine model predictions to validate their biological evolutionary relationships, requiring the integration of original single-cell data with model outputs.
The spatiotemporal complexity of the single-cell data and the large number of connections between different samples often make this task multifaceted.

To tackle the above tasks, we propose \system, a visual analytics system with a computational model that assists biologists in predicting and exploring cross-sample cell developmental trajectories. 
Firstly, we construct a GNN-based model that comprehensively captures cross-sample cellular developmental features, including cellular spatial distributions, gene expression profiles, and cross-sample dynamics, to predict potential developmental trajectories for multi-sample \scrna datasets. 
Secondly, \system enables users to explore predicted trajectories and validate their biological evolutionary relationships through interactive explorations, where users can analyze cellular spatial distribution patterns, developmental directions, and the functions of expressed genes along these trajectories. 
Finally, we quantitatively evaluated our proposed model on three multi-sample \scrna datasets, demonstrating its effectiveness in predicting cross-sample cellular developmental trajectories.
We also conducted two case studies and expert interviews to validate the effectiveness of our system qualitatively.

In conclusion, the contributions of our work are: 
\begin{itemize}
	\item Problem formulation and design requirements for defining and analyzing cross-sample cell developmental trajectories.
	\item A GNN-based model to predict cross-sample trajectories by capturing cross-sample cellular developmental features.
	\item \system, a visual analytics system for exploring and identifying biological evolutionary relationships of the predicted trajectories.
	\item Quantitative evaluation of our model's performance and qualitative validation of our system's effectiveness through two case studies and interviews with eight experts.
\end{itemize}

\begin{figure*}[t]
	\centering
	\includegraphics[width=\linewidth]{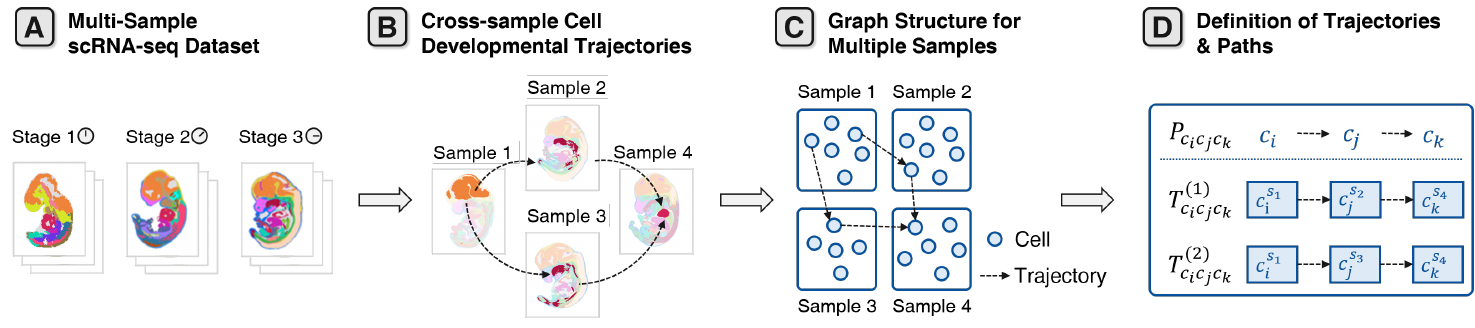}
	\caption{
		We define the evolutionary relationships between cells across samples as cross-sample cell developmental trajectories. 
		Next, we abstract the dataset into a graph, with each sample as a subgraph forming the whole graph, and refer to sequences of cross-subgraph links as paths. 
	}
	\label{fig2}
\end{figure*}

\section{Related Work}

In this section, we summarize the prior work related to our research on cell developmental trajectory inference tools, single-cell data visualization, and spatiotemporal data visualization.

\subsection{Cell Developmental Trajectory Inference Tools}

Methods for constructing cell developmental trajectories based on \scrna data, known as ``pseudo-time'' analysis, temporally order cells within individual samples by comparing gene expression patterns and assign pseudo-time values to individual cells \cite{trajectory2024pan}. 
Tools like PAGA \cite{paga2019wolf} build graph-based trajectories, with nodes as cell clusters and edges as developmental paths. 
DTFLOW \cite{dtflow2021wei} uses feature decomposition and reverse search, while VeloViz \cite{veloviz2022atta} integrates \scrna data and RNA velocity data \cite{rnavelocity2018manno} . 
However, these tools limit in cross-sample trajectory construction, as they primarily rely on single-sample analysis and fail to capture multi-dimensional features like spatial dynamics and gene expression across samples. 
Consequently, there is a need for a specialized tool to construct and analyzing complex cross-sample developmental trajectories.

\subsection{Single-cell Data Visualization}

Single-cell data, primarily \scrna, captures individual cell characteristics \cite{singlecell2022bo}, often integrated with spatial transcriptomics and protein expression data for comprehensive cellular state analysis.
Current \scrna data analytical tools like Seurat \cite{seurat2024hao} and CellRanger\cite{cellranger2025genomics} provide visualization for dimensionality reduction, cell clustering, and differential gene expression analysis. 
In addition, some tools explain complex cell dynamics through specialized visualizations. For example, Cellchat \cite{cellchat2025jin} maps intercellular communication networks via ligand-receptor interaction matrices, and PAGA \cite{paga2019wolf} visualizes developmental trajectories as graph structures.
Visual analytics tools enhance single-cell data exploration by integrating additional metadata, such as temporal dynamics \cite{multeesum2010meyer}, spatial contexts \cite{visinity2023warchol}, and multi-modal datasets \cite{cellenium2023jahn, vitessce2025keller}. 
For instance, MulteeSum \cite{multeesum2010meyer} supports comparisons across multiple samples over time series. 
Visinity \cite{visinity2023warchol} incorporates spatial information to help users understand the positional relationships of cells within tissues. 
Additionally, some tools leverage user interactions to improve data exploration accuracy. Polyphony \cite{polyphony2023cheng} uses user-selected anchors to refine dataset alignment. LineageD \cite{lineaged2022hong} assists users in constructing cell lineages, enhancing the precision of lineage mapping tasks.
However, critical challenges persist in cross-sample trajectory analysis, as current tools struggle to integrate and track cellular state changes across multiple samples and time points, underscoring the need for advanced tools to address these limitations.

\subsection{Spatiotemporal Data Visualization}

Spatiotemporal data encompasses entities with inherent spatial and temporal dimensions \cite{spatiotemporal2012han}, enabling the study of dynamic systems like biological trajectories \cite{eco2024fouqueau}, environmental transitions \cite{greenspace2021rahaman}, and urban dynamics \cite{urban2024misty} .
Spatiotemporal visualization tools typically address domain-specific needs.
For geospatial-temporal phenomena, GeoChron \cite{geochron2024deng} presents the evolution of geographic entities over time with the combination of maps and narrative timelines. 
BNVA \cite{bus2021weng} optimizes bus routes by discovering passenger flow patterns and comparing characteristics of different routes. 
UcVe \cite{ucve2023yu} assists users in comparing multiple visualized geographic units within spatiotemporal space.
Additionally, spatiotemporal visualization is also used to explore relationships between entities and model relational dynamics. 
MoReVis \cite{morevis2024g} visualized moving entities to facilitate analysis applications in fields such as climate science. 
SpreadLine \cite{spreadline2025kuo} focuses on dynamically illustrating influence within egocentric networks by employing a storyline-based design to represent entities with their evolving relationships. 
Wallner \etal \cite{lol2023wallner} utilized storyline techniques to integrate the development of entity relationships over time and geographical trajectories in MOBA games to facilitate game replay analysis.
Spatiotemporal data visualization can also be applied to analyze and compare cell spatial distributions across multiple biology samples, thereby facilitating the exploration of how cellular patterns dynamically change and develop over both time and space.
However, current spatiotemporal visualization methods have limitations in effectively integrating discrete biological events with continuous spatial dynamics, which calls for a visual analytics system that integrates multi-dimensional data and supports interactive temporal-spatial linking to enable biologists to validate the biological evolutionary relationship of cross-sample cell developmental trajectories more accurately.

\section{Design}

In this section, we introduce the problem formulation of our research, as Fig.\ref{fig2} shows. We also detail the collaboration process we adopted with experts during the design process as well as the specific analysis tasks we identified to guide our system design effectively.

\subsection{Terminology Definitions}
Single-cell RNA sequencing (\textbf{scRNA-seq}) is a widely used technique for detecting and quantifies messenger RNA (mRNA) \cite{mrna2009tang, celseq22016hashimshony}, the transcriptional product of DNA \cite{mrna1970crick}, within individual cells, thereby helping biologists understand gene expression patterns \cite{tutorial2021andrew}. 
\textbf{ScRNA-seq datasets} typically consist of multiple biological samples \Fig{\ref{fig3}}{A}, representing the gene expression profile of a single cell \cite{dataset2023sount}.
Common analytical methods include clustering \cite{cluster2024xie}, cell-cell communication analysis \cite{communication2018kumar}, and developmental trajectory inference \cite{paga2019wolf}.
A \textbf{cell developmental trajectory} refers to the path that a cell population follows from an undifferentiated state to a differentiated one \cite{differentiate2021staseen} \Fig{\ref{fig3}}{B}. 
Since real-time tracking of all cells across continuous time points is experimentally impractical, current trajectory inference methods simulate a ``pseudo-temporal'' timeline by sorting cells based on gene expression similarity \cite{dtflow2021wei}, allowing researchers to infer the developmental sequence and identify key cell types involved in biological processes \cite{geneexprerssion2022wei}.

\subsection{Problem Formulation} \label{sec3.1}
Based on the above definitions, we model cell developmental trajectories as graphs from multi-sample \scrna datasets, with nodes as cell types and edges as cell developmental trajectories \Fig{\ref{fig3}}{C}.
Cross-sample developmental trajectories for specific cell populations were defined and unified into cross-sample edges \Fig{\ref{fig3}}{D}.

\textbf{Graph Construction.} 
Our \scrna datasets consist of multiple samples representing different developmental stages of a biological subject, with each sample tied to a specific stage, though stages may include multiple samples.
Collaborative discussions with biologists, revealed that developmental trajectories within samples are often abstracted as graph structures \cite{dtflow2021wei}.
Inspired by this, we formally built a global graph $G=(V, E)$ to depict a complete cell developmental trajectory. The vertex set $V$ can be represented as:
\[
V = \bigcup_{s=1}^k V^{(s)}, \ \text{where} \ V^{(s)} = \{v^{(s)}_i \} \text{.}
\]
Here a vertex $v_{i}^{s}$ represents cell population $i$ (identified by annotated cell names) in sample $s$. And the edge set $E$ can be defined as:
\[
E=\{e \mid e = (v_{i}^{(p)},v_{j}^{(q)}) \ ,\ \text{stage}(p) \neq \text{stage}(q) \} \text{,}
\]
where $\text{stage}(p)$ denotes sample $p$'s developmental stage, ensuring edges only connect cell populations from temporally distinct stages.

\textbf{Path Definition.} 
To comprehensively characterize cellular differentiation dynamics across samples and developmental stages \cite{multicellular2020fisher}, we formally defined a developmental trajectory as a temporally ordered sequence of cell type progressions, represented as: 
$$T_{c_{1}c_{2} \dots c_{m+1}}=[c_{1}^{s_{a}} \overset{e_1}{\rightarrow} c_{2}^{s_{b}} \overset{e_2}{\rightarrow} \dots \overset{e_m}{\rightarrow} c_{m+1}^{s_{c}}] \text{,}$$ 
where $c_i^{s}$ represents a cell type $i$ in sample $s$, and $e_{i}$ represents a developmental trajectory between cell types $c_{i}$ and $c_{i+1}$. 
As several trajectories may share the same cell-type sequences but originate from different samples, we defined a cross-sample cell developmental path as a collection of trajectories with identical same cell-type sequences across different samples (\eg $c_{1}^{s_{1}} {\rightarrow} c_{i_{2}}^{s_{2}} {\rightarrow} c_{i_{3}}^{s_{3}}$ and $c_{i_{1}}^{s_{1}} {\rightarrow} c_{i_{2}}^{s_{4}} {\rightarrow} c_{i_{3}}^{s_{5}}$) : 
$$ P_{c_{1}c_{2}c_{3}} = \{ T_{c_{1}c_{2}c_{3}}^{(1)}, T_{c_{1}c_{2}c_{3}}^{(2)}, \cdots, T_{c_{1}c_{2}c_{3}}^{(n)} \} \text{,}$$ 
where $n$ represents the frequency of the paths, \ie, the number of distinct sample combinations supporting the same cell type sequence.

\subsection{Design Process} \label{sec:section3.2}

We closely collaborated with six biologists (\Ex{1}-\Ex{6}) with varying years of experience across specialized fields. Specifically, \Ex{1} and \Ex{2} specialize in bioinformatics, with 7 and 4 years of experience, respectively.
\Ex{3} has 5 years of experience in microbiology, and \Ex{4} brings 5 years of expertise in cell biology.
\Ex{5} and \Ex{6} are medical researchers with experience in using bioinformatics tools, possessing 4 and 5 years of research experience.
All participants have experience using tools to analyze \scrna data by conducting cell developmental trajectories. 
We conducted at least two rounds of systematic interviews with each expert, obtaining insights into their workflows and the challenges they encountered in studying cell developmental trajectories. 
This collaboration ensures that our systems are aligned with the needs of the biological field, while the diversity of expert backgrounds helps mitigate bias arising from individual perspectives.

\subsection{Analysis Tasks}

Based on interviews with six experienced biologists, we have summarized five key analysis tasks (\textbf{T1} - \textbf{T5}) that users need to explore when investigating cross-sample cell developmental trajectories.

\begin{itemize}[leftmargin=1.8em]
	\item[\textbf{T1}] \textbf{Select a specific cell type for analysis based on its temporal occurrence.}
	Domain experts emphasized that selecting a core cell type for cross-sample developmental trajectory analysis should prioritize cell types persisting consistently across multiple developmental stages. Their consistent presence across stages suggests they play critical roles in biological development, thereby determining them as key candidates for further investigation.
	
	\item[\textbf{T2}] \textbf{Identify high-frequency predicted cell developmental paths related to the selected cell type.}
	After selecting a core cell type, biologists aim to identify high-frequency developmental paths connecting it to other cells undergoing critical dynamics, such as those that emerge, disappear, or experience quantitative changes during key developmental stages. 
	As \Ex{2} noted, such cells are central to driving biological processes, as their temporal dynamics often reflect pivotal mechanisms: \textit{``Cells that emerge or disappear during key developmental processes typically play important roles in shaping the underlying mechanisms.''}
	
	\item[\textbf{T3}]  \textbf{Assess cell developmental paths based on cellular spatial continuity.} 
	Biologists assess the spatial continuity and directional consistency of cellular populations along developmental trajectories to validate their evolutionary relationships and dynamic behaviors. 
	As noted by \Ex{4}, \textit{``Trajectories with biological evolutionary relationships typically maintain spatial continuity between early and subsequent cell populations.''} 
	By integrating these spatial and directional features, biologists can enhance the understanding of the biological evolutionary relationships of these paths.
	
	\item[\textbf{T4}]  \textbf{Examine original samples as context information.} 
	For validated developmental paths, each of them typically encompasses multiple trajectories. 
	Due to sample-level differences, such as developmental stages or cell scale differences, developmental trajectories may exhibit differences in gene expression. 
	Therefore, experts need to select multiple trajectories within each path to determine if they exhibit consistent or divergent patterns.
	
	\item[\textbf{T5}] \textbf{Identifying differentially expressed genes in specific trajectories and their biological functions.}
	Experts select multiple trajectories within each path to examine gene functions separately, as different trajectories may exhibit distinct gene expression patterns.
	Specifically, in bioinformatics practices, identifying differentially expressed genes (DEGs) is a pivotal approach to delineating the genetic foundations of cell developmental trajectories \cite{deg2024yin}. 
	To achieve this goal, biologists typically utilize statistical methods \cite{go2024candia} and gene annotation databases \cite{go2023the} to uncover the biological processes in which these DEGs are involved.
	
\end{itemize}

\section{Data Processing}

\begin{figure}[t]
	\centering
	\includegraphics[width=\linewidth]{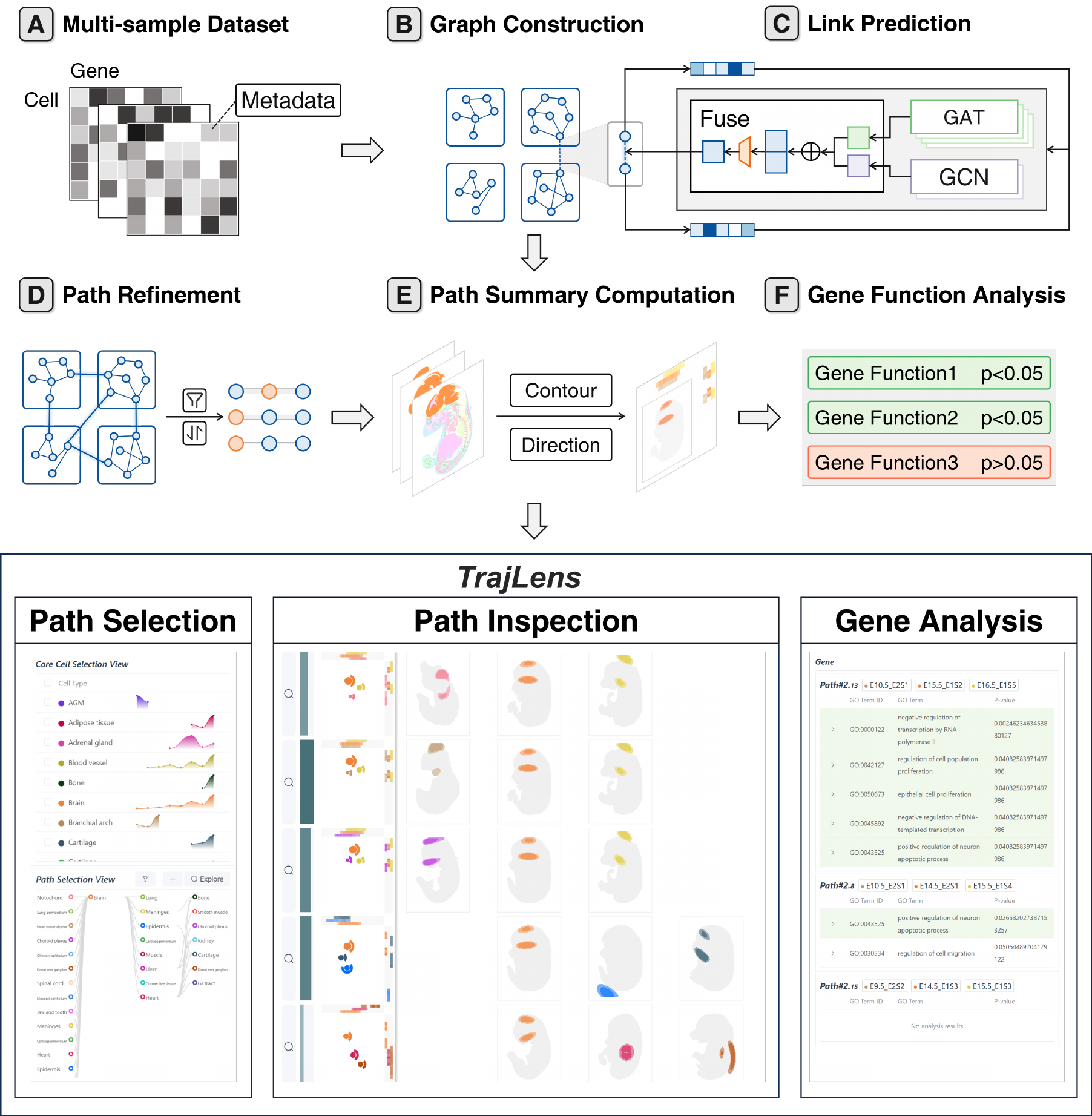}
	\caption{
		The data processing and analyzing workflow of \system is as follows: 
		Each sample of our multi-sample \scrna datasets is represented as a matrix containing cell and gene information with single-cell-level metadata describing cellular characteristics. 
		We then abstracted the dataset into a unified graph structure, where each sample forms a subgraph collectively integrating into a large graph. 
		Next, we conducted a GNN-based model to predict links between subgraphs, \ie, cross-sample cell developmental trajectories. 
		Then, we identified high-frequency developmental paths composed of multiple cell types for user selection.
		We computed a summary overview for each selected developmental path and provided gene function analysis along specific paths.
	}
	\label{fig3}
\end{figure}

In this section, we introduce our data processing workflow, which analyzes multi-sample \scrna datasets and predicts cross-sample cell developmental paths (Fig.\ref{fig3}). 
Firstly, we collected three multi-sample \scrna datasets, preprocessed them following biological practices, and constructed them into graphs (\Fig{\ref{fig3}}{A}).
Secondly, we implemented a GNN-based model and conducted training and prediction phases (\Fig{\ref{fig3}}{B, C}).
Subsequently, we filtered predicted edges according to specific biological criteria and identified high-frequency paths (\Fig{\ref{fig3}}{D}).
Finally, we provided a comprehensive summary of paths, including path overviews and gene function analysis (\Fig{\ref{fig3}}{E, F}).
Through these steps, \system provides a complete workflow for processing and analyzing multi-sample \scrna data, enabling biologists to better understand the dynamic changes in cell development.

\subsection{Dataset}\label{sec:4.1}

\textbf{Datasets Description.} 
We collected a multi-sample \scrna dataset for the following analysis process, referred to as Mouse-Embryo Dataset, whose detailed descriptions are provided in Supplement Materials.
Each sample in the dataset contains a gene expression matrix and single-cell-level metadata . 
In the gene expression matrix, each row corresponds to a unique single cell, with each column representing a gene type and the values indicating the expression levels of a specific gene within the cell. 
Additionally, the single-cell-level metadata includes spatial coordinates of every single cell (\ie, $x$ and $y$ coordinates in a two-dimensional plane) for describing its extract position within the sample. 
Furthermore, cells are annotated by their cell types, such as neuronal progenitor cells, endothelial cells, and many others. 

\textbf{Dataset Preprocessing.} 
For each dataset, we normalized the expression values of each sample \cite{preprocess2017zheng} and employed the widely-used bioinformatics tool Harmony \cite{harmony2019ilya} to mitigate technical variations across multiple samples.
For each sample, we constructed a graph-based representation $G_{sub}=(V, E_{none})$ to model cellular relationships, where nodes $v \in V$ correspond to annotated cell types. 
Notably, edges $E_{none}$ in each graph were not predefined, as we will construct a model to predict them. 
We also stored metadata for each node, including a set of differentially expressed genes and spatial coordinates of all cells within the node, which provides critical biological context for downstream analysis.
Finally, we construct a global graph $G=\{ G_{sub}^{(1)}, G_{sub}^{(2)}, \cdots, G_{sub}^{(s)}\}$ composed of multiple sample-specific subgraphs $G_{sub}$.

\subsection{Model Setup}\label{sec4.2}
\textbf{Adoption Rationale.}
Existing methods, such as PAGA and linear models, fail to adequately capture implicit semantic relationships \cite{model2020denis}, such as node appearance/disappearance events and cell gene expression profiles. Consequently, they face challenges in effectively performing cross-sample cell developmental trajectory prediction. Consequently, they limit in predicting cross-sample cell developmental trajectories effectively. 
In contrast, the Graph neural network (GNN) architectures \cite{gnn2009franco} excels at preserving both node-level features and complex topological relationships \cite{gnn2023han}, making them particularly suitable for representing dynamic cellular development.
Consequently, we conducted a GNN-based model that effectively models the spatiotemporal dynamic graph structure of our multi-stage cellular development dataset.

\textbf{Training Data Construction.} 
We constructed a graph as training and testing data based on Dataset 1 using the PAGA algorithm, where nodes represent cell populations \cite{leiden2019traag} and edges denote connections inferred by PAGA. 
Moreover, as the model must not only capture the graph structure, \ie, the nodes and edges generated above, but also characterize node features, including gene expression profiles and cell distribution patterns, we define features for each node. 
Specifically, we first identified the top 20 differentially expressed genes (DEGs) for each cell population through the Wilcoxon rank-sum test \cite{scanpy2018wolf}. 
This choice of 20 DEGs is a commonly used approach and has been demonstrated to distinguish between cell types effectively \cite{ideas2022zhang}.
We first embedded these DEGs into 1024-dimensional semantic embeddings using the pre-trained model BioLinkBERT \cite{linkbert2022yasunaga}. 
Simultaneously, we performed Principal Component Analysis (PCA) \cite{pca1901pearson} on the spatial coordinates of cell populations, which can be represented as an $N \times 2$ matrix containing $(x,y)$ coordinates for $N$ cells, and reduced them to 1024-dimensional vectors enabling comprehensive analysis of cellular features and differentiation trajectories across samples.

\textbf{Model Construction.} 
Specifically, our model integrated two components: a two-layer Graph Attention Network (GAT) \cite{gat2018petar} module with multi-head attention for modelling inter-cell interactions and a two-layer Graph Convolutional Network (GCN) \cite{gcn2017kipf} module for global topological edge propagation. 
Through residual connections, we fused local attention patterns and global structural information, enabling complementary optimization of both modules. 
The model inputs a graph where each node has a 2048-dimensional feature embedding and outputs probabilistic predictions of edges. 
By testing different thresholds ranging from 0.5 to 0.9 on the test dataset, we found that the highest precision was achieved at a threshold of 0.75. Consequently, edges with probabilities greater than 0.75 are treated as existing, thereby constructing the predicted cell developmental trajectories.

\textbf{Model Training and Predicting.} 
To enhance model robustness during training, we generated negative sample edges in each training epoch and implemented batch shuffling to prevent overfitting. After model training, we used the graphs constructed in Section \ref{sec:4.1} that only consist of nodes without edges as input data and employed the optimized weight parameters from the training process to predict edges.
All models are trained and evaluated using PyTorch on a 48-GB NVIDIA RTX 4090 GPU, for 100 epochs with a batch size of 8. The model has approximately 1,000,000 parameters and is optimized with AdamW \cite{adam2019losh} at an initial learning rate of $1 \times 10 ^{-5}$.

\subsection{Cross-sample Developmental Path Refinement} \label{sec4.3}

Based on trajectory predictions generated by our proposed GNN-based model, we integrated subgraphs for each sample with predicted edges to construct a comprehensive directed graph $G_{pred}=(V, E_{pred})$. 
Each node $v \in V$ contains metadata such as cell type name, sample origin, temporal developmental stage, differentially expressed genes, and spatial coordinates of each single cell. 
The directed edges $e \in E_{pred}$ symbolize predicted developmental trajectories based on their origin $src$ and terminus $dst$, which were filtered out under biological and temporal constraints to ensure that they represent cross-sample developmental trajectories: 
edges between different cell types $type(e_{src}) \neq type(e_{dst})$ and those directed from earlier to later stages $stage(e_{src}) < stage(e_{dst})$. 
Based on filtered edges, we merged redundant edges with identical source-target pairs by counting their occurrences, defining the merged edge weights as the count of original edges:
$$w(u \to v) = \lvert \{ e \in E_{filtered} \mid src(e)=u \wedge dst(e)=v\} \rvert \text{,}$$
where $w$ represents the weights of the merged edge, quantifying the total count of original edges that were combined.

Given the high density $\rho=\frac{\lvert E \rvert}{\lvert V \rvert(\lvert V \rvert-1)}\approx50\%$ of the initial graph which was biologically suboptimal because the densely connected graph tends to contain substantial noise edges while obscuring critical biological signals \cite{network2004barabasi}. 
To mitigate this, we retained only the top 15\% most frequent paths, reducing the analysis complexity. 
The threshold selection aligned with typical biological practices in single-cell data analysis, retaining 5\% to 15\% of high-confidence features for core analysis \cite{threshold2019yip}. 
For the user-selected key cell type, we conducted a breadth-first search (BFS) on $G_{pred}$ to trace its ancestral and descendant cell populations along developmental trajectories. 
We then constructed a hierarchical tree rooted at the key cell type, with each node recording its topological distance from the root and its relationship as either an ancestor or descendant. 
Multiple cell types in the tree constitute a cross-sample cell developmental path.

\subsection{Path Summary Computation} \label{sec4.4}

To help biologists understand the spatial distribution characteristics of cell developmental trajectories more intuitively, we generated detailed contour maps for each cell population along the developmental path to visualize their global spatial distribution patterns. We also provided a similarity metric between contour maps to identify cell populations with similar spatial distributions. Additionally, we derived cell developmental directions from spatial density gradients to reveal potential cellular evolutionary trends during cell development.

\textbf{Contour map computation.} 
With the spatial range of a specific sample, we created a $ 100 \times 100$ grid and used Gaussian Kernel Density Estimation to compute the probability density distribution of grid points based on the spatial distribution of cells within the sample. 
Then, we generated polygonal contours for predefined thresholds using the Marching Squares algorithm \cite{marchingsquares1987lorensen} and simplified these contours using the Douglas-Peucker algorithm \cite{douglasalgorithm2011douglas}.
The contour coordinates were projected onto the $x$- and $y$-axis and binned into 100 histogram intervals weighted by their occurrence frequency on these axes.
To provide users with a spatial reference for cell contours, we utilized the Alpha Shape algorithm\cite{alphashape1983edel} to generate geometric shapes that capture the biological boundaries of biology samples and extract the $x$ and $y$ coordinates of these shapes. 

\textbf{Contour similarity computation.}
When calculating the similarity of given two distribution contours $C_1$ and $C_2$ from cell type pairs, we bidirectionally computed the minimum point-wise distances in both directions: $d_{C_{1} \rightarrow C_{2}}$ and $d_{C{2}} \rightarrow d_{C_{1}}$.
For each point $p_i$ in contour $C_1$, we computed the minimum distance to each point in contour $C_2$: $d_{p_i \rightarrow C_2} = \min_{q_j\in C_2}{\left \| p_i - q_j \right \| } $. The overall distance $d_{C_1 \rightarrow C_2}$ is then obtained by taking the mean of all point-wise minima:
$$d_{C_1 \rightarrow C_2} = \frac{1}{\lvert C_1 \rvert}\sum_{i=1}^{\lvert C_1 \rvert}{d_{p_i \rightarrow C_2}}\text{.}$$
The symmetric spatial overlap  $D_{sym}$ is then defined as the reciprocal of the mean bidirectional distances: 
$$D_{sym} = \frac{2}{d_{c_2\rightarrow c_1}+d_{c_1\rightarrow c_2}}\text{.}$$
Higher $D_{sym}$ values indicate more significant similarity between the two contours, while lower values reflect distribution differences. This metric quantitatively compares spatial characteristics across cell types.

\textbf{Cell developmental direction computation.}
Subsequently, we applied Principal Component Analysis (PCA) to identify the primary developmental directions of the contour map for each cell. Specifically, the input to PCA consisted of the set of two-dimensional coordinates $(x, y)$ of each cell and the output provided the primary developmental direction of the cell's contour map, which was quantified with a range $[0, 2\pi]$ and calculate the stability ratio as follows:
$$R_{std} = \sqrt{\frac{\lambda_2}{\lambda_1 + \lambda_2}},$$
where $\lambda_1$ and $\lambda_2$ represent the eigenvalues of the primary and secondary components, respectively.
The value $R_{std}$ close to 0 indicates high stability in the developmental direction of a specific cell.

\subsection{Gene Function Analysis}

We analyzed gene functions of cell developmental trajectories using Gene Set Enrichment Analysis (GSEA) \cite{gesa2005Subramanian} combined with the Gene Ontology (GO) database \cite{go2000ash, go2023the}.
Specifically, we constructed a hierarchical network of GO terms based on the GO database using GOATOOLS \cite{goea2018klop}, mapped genes based on their mapped genes to their functional terms via mouse Gene Association File (GAF) annotations\cite{godb2024baldarelli}.
For a specific developmental path, we used a statistical test method \cite{goea2018klop} to assess the enrichment of GO terms in the target gene set (highly expressed genes of specific cell types) relative to the background gene set (highly expressed genes of all samples), identifying overrepresented biological functions of these genes based on the GAF file.

\section{System}

In this section, we introduce the visual designs of \system. 
Within our analytical workflow, users first select core cell types based on temporal occurrences and population dynamics of cells in the \celltable (\textbf{T1}). 
Subsequently, \system presents other cell types predicted to have potential developmental trajectories related to the core cell types in the \pathselection, where users can further select multiple cell combinations as candidate paths for comparative analysis (\textbf{T2}). 
Then, users can evaluate the biological evolutionary relationships of these paths through multi-row visualization with cellular spatial distribution in the \pathprojection (\textbf{T3}). 
Additionally, they can delve into the trajectories constituting specific paths in the \trajectorycomparison (\textbf{T4}) and utilize the \genefunction to analyze the enriched gene functions presented in these trajectories (\textbf{T5}).

\begin{figure*}[ht]
	\centering
	\includegraphics[width=\linewidth]{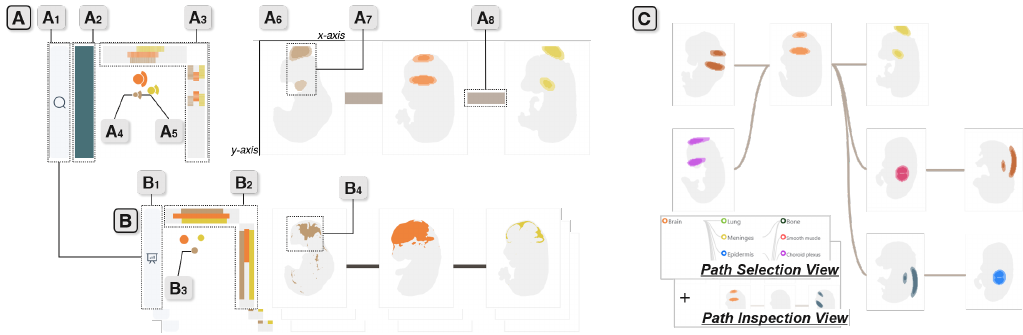}
	\caption{
		The design of the multi-row visualization in Section \ref{sec5.2}: 
		(A) A row in the \pathcomparison represents a cell developmental path defined in Section \ref{sec3.1}.
		(B) The \trajectorycomparison supports the detailed inspection of the developmental trajectories that constitute the path in (A).
		(C) The design alternatives of this view integrates the \pathselection and \pathcomparison.
	}
	\label{fig4}
	\vspace{-1em}
	\mbox{}
\end{figure*}

\subsection{Cell Developmental Path Selection}

Users determine paths for comparative analysis in two steps: first, selecting core cell types based on temporal occurrences and population dynamics (\textbf{T1}); then, selecting cell types predicted to have developmental relationships with the selected cells, forming multiple cell combinations as candidate paths for exploration (\textbf{T2}).

\subsubsection{Core Cell Selection View}

To help users select core cell types based on their temporal occurrences and quantitative changes during biological development, we propose the \celltable \Fig{\ref{fig1}}{A}. 
The \celltable combines a tabular layout with line charts, where each row displays a cell type's name along with its quantitative changes and occurrence points across developmental stages.
The horizontal axis of each row's line chart represents different developmental stages, and the vertical axis represents the quantity of cells.
In addition, each cell's color in the \celltable corresponds to the standardized color-encoding defined in the original dataset, where cell types are consistently assigned specific colors based on their biological classification (\eg muscle cells in red, bone cells in green) to maintain consistency with prior studies. \footnote{\href{https://db.cngb.org/stomics/mosta/spatial/}{https://db.cngb.org/stomics/mosta/spatial/}}
Users can identify cells of interest by examining their presence and quantity changes over developmental stages. 
Cells existing along the shown developmental stages are typically selected as cells of interest for further investigation.

\subsubsection{Path Selection View} \label{sec:5.1.2}

Upon selecting a core cell type, users aim to explore sequences involving more than two types of cells relevant to the selected cell, which is defined to cell developmental paths in Section \ref{sec3.1}.

\textbf{Description.} 
When selecting candidate cell developmental paths, users typically prioritize those with high frequency. 
To address this requirement, we design the \pathselection \Fig{\ref{fig1}}{B}, which supports filtering and visualizing high-frequency cell developmental paths while enabling intuitive selection and analysis of paths of interest. 
Given that the data involving the selected cell of interest and its neighboring cells forms a hierarchical structure (Section \ref{sec4.3}), where multiple developmental paths share common cell types and diverge at these shared cells, we adopt a hierarchical tree visualization for \pathselection to present the relationships between paths and their shared nodes explicitly.
The view places the core cell types as the root node \Subfig{\ref{fig1}}{B}{1}, with adjacent columns presenting cells having predicted paths leading to it. 
Nodes in each column represent individual cell types, and links connect nodes across adjacent columns, indicating predicted cell developmental paths.
To emphasize high-frequency paths, users can click the ``Filter'' button to exclude paths with frequencies less than a specific threshold, thereby focusing on more representative developmental paths. 
Additionally, the vertical order of nodes within each column is determined by the number of paths connected to them, which aligns with the users' requirement to explore cells that play pivotal roles in diverse developmental processes.

\textbf{Interaction.} 
When hovering over a node in the \pathselection, users can access a line chart about the temporal occurrence and quantity changes of the corresponding cell type through a tooltip \Subfig{\ref{fig1}}{B}{2}. 
Leveraging the tooltip, users can select multiple cells that emerge or disappear during development or show substantial quantity changes to form a developmental path. 
\system will then automatically check for connections between the selected nodes. 
If the path is valid (\ie, all consecutive nodes are connected), it will be incorporated into the list for inspection in the following process.
After selecting all desired paths, users can click the ``Explore'' button, and our system will proceed with further in-depth analysis of these paths.

\subsection{Cell Developmental Path Inspection} \label{sec5.2}

Biologists need to validate the biological evolutionary relationships underlying the selected developmental paths in Section \ref{sec:5.1.2} by assessing cellular spatial distributions across them (\textbf{T3}).
To facilitate this task, we propose the \pathcomparison, featuring multi-row visualization for comparing spatial overlap and development trends among cell populations from multiple samples, along with an aggregated overview for intuitive comparison.
Additionally, to refine the analysis of cell developmental trajectories constituting specific paths (\textbf{T4}), we introduce the \trajectorycomparison for analyzing trajectories constituting specific paths, supporting the following gene function analysis. 

\subsubsection{Path Inspection View}

The \pathcomparison consists of multi-row visualizations, where each row represents a selected developmental path, with the summary visualization for each row offering an overview of the path.
Additionally, the similarity visualization is above the multi-row visualization to illustrate the similarity of node contour maps within a column.

\textbf{Description.}
The \pathcomparison employs a node-link diagram to represent cells within the selected paths with cellular spatial distribution \Fig{\ref{fig4}}{A}. 
Each row of this view corresponds to a selected path in Section \ref{sec:5.1.2} \Fig{\ref{fig4}}{A}, where nodes correspond to cell types in the path \Subfig{\ref{fig4}}{A}{6} and edges depict predicted developmental trajectories with their color intensity and height encoding the similarity between consecutive cells' spatial distribution \Subfig{\ref{fig4}}{A}{8}. 
We visualize the overall distribution of each node through layered contour maps: the outer contour represents 98\% core density coverage,  while the inner contour shows 94\% density \Subfig{\ref{fig4}}{A}{7}, which enables researchers to simultaneously capture the most concentrated core and the broader distribution of cell populations.
The color intensity of these contours is scaled to reflect the corresponding density coverage, providing a visual hierarchy of cell distribution. 
All sequences are spatially aligned based on the user-selected core cell to ensure cross-row consistency.
To address the limitation that the \pathcomparison alone cannot provide an overview of a sequence's overall characteristics, we introduce the summary visualization of each row.
Within each summary visualization, central dots represent the centroid of each cell population with their radius indicating the average cell count in samples \Subfig{\ref{fig4}}{A}{4}. 
Fan-shaped arcs around the dots encode the primary developmental direction of the cell population through the central angle, and angular span reflects their directional variance computed in Section \ref{sec4.4} \Subfig{\ref{fig4}}{A}{5}.
Additionally, we design a multi-projection visualization \Subfig{\ref{fig4}}{A}{3} that orthogonally projects the spatial distributions of each node onto the $x$- and $y$-axis and then stacks them sequentially. 
The innermost projections represent the leftmost node in the path, and subsequent projections expand outward to indicate the nodes' order in the path.
Each row also features a bar chart where the width and color intensity indicates path frequency \Subfig{\ref{fig4}}{A}{2} and a button to the left to the view opens a \trajectorycomparison in a new tab upon selecting and leads to the following analysis\Subfig{\ref{fig4}}{A}{1}.

\textbf{Interaction.}
Upon selecting a row by clicking the leftmost button \Subfig{\ref{fig4}}{A}{1}, we present the similarity visualization for each node of the selected path and other nodes in the corresponding column by bar charts above the \pathcomparison \Subfig{\ref{fig1}}{C}{1}. 
The vertical order of the bars matches the vertical order of the nodes in the column, facilitating users to identify and select additional paths similar to the selected ones.

\textbf{Design alternative.}
During the iterative design process, we refined \pathcomparison based on user feedback. 
Initially, we integrated the \pathselection and the \pathcomparison into a hierarchical tree visualization with contour maps for each node that presented cell type relationships and spatial distributions \Fig{\ref{fig4}}{C}. 
However, while effective for exploring cell relationships, this approach is limited in a detailed comparison of divergent branches. 
Furthermore, a node in this view may represent the same cell type from multiple paths, potentially originating from different sample sets, complicating the clear representation of a single node.
Therefore, we separated the hierarchical tree from comparison views, allowing clearer comparison of different paths in the \pathcomparison and preserved the hierarchy-based exploration of cellular relationships in the \pathselection.

\subsubsection{Trajectory Inspection View}

Selecting a row in \pathcomparison opens a new tab displaying \trajectorycomparison, which, like \pathcomparison, includes multi-view and summary visualization components. Each node in the \trajectorycomparison \Fig{\ref{fig4}}{B} corresponds to the cell distribution map of a single sample \Subfig{\ref{fig4}}{B}{4}.

\textbf{Description.}
For summary visualization, we marked the centroids of cell populations with dots, where the dot positions represent spatial centroids and their sizes encode population counts \Subfig{\ref{fig4}}{B}{3}. Additionally, Additionally, we projected the spatial distributions of cell populations onto the $x$- and $y$-axis as density profiles \Subfig{\ref{fig4}}{B}{2}. 

\textbf{Interaction.}
By clicking the button on the left side of the specific row \Subfig{\ref{fig4}}{B}{1}, \system will automatically conducts gene function analysis based on the gene set of this trajectory and present the analysis result in the \genefunction in a table format.

\section{Case Study}

We conducted two case studies by recording the exploring process of two interviewed experts, \Ex{1} and \Ex{5}, using our system, and then summarized their insights to demonstrate the practicality of our system.

\subsection{Case Study I: Exploring Cross-sample Developmental Paths in Mouse Embryonic Development Dataset}

As an expert in bioinformatics, \Ex{1} used \system to explore the cross-sample cell development paths related to the brain cell type and investigate the gene functions underlying these paths.

\textbf{Selecting the Core Cell Type (T1).} 
Among the line charts of multiple cell types in \celltable \Fig{\ref{fig1}}{A}, \Ex{1} identified distinct patterns of cellular dynamics throughout developmental stages in Dataset 1. 
Many cell types exhibited transient emergence, appearing only at specific developmental stages (\eg AGM cells transiently emerged at the first two embryonic days to generate stem cells), making them less informative for analysis. 
In contrast, the line chart depicting the emergence of brain cells exist throughout all developmental stages \Subfig{\ref{fig1}}{A}{1} in the \celltable suggesting they may play a crucial role in nervous system formation. 
Consequently, \Ex{1} selected the brain cells as the core cell type for in-depth exploration.

\textbf{Selecting Cross-sample Cell Developmental Paths (T2).}
After selecting the brain cells as the core cell type, \pathselection presented all associated cell types hierarchically, including direct precursors and successors as well as indirectly linked cells \Fig{\ref{fig1}}{B}.
\Ex{1} filtered out paths with frequencies less than 40 to reduce analysis complexity, as low-frequency paths were less likely to be involved in critical developmental processes.
Based on their ranking in the column and temporal occurrence and quantity changes shown in tooltip line charts \Subfig{\ref{fig4}}{B}{2}, \Ex{1} selected cells such as the Notochord, Choroid plexus, and Dorsal root ganglion, \etc
Their tooltip line charts spanning only specific stages suggest these cells may differentiate into other cell types or originate from them and are more likely to participate in brain-related developmental processes.
Some highly ranked cells in their columns, such as Cartilage, were not selected due to their persistent presence without significant changes.
This approach focused on dynamically changing cell types with potential critical functions.

\textbf{Inspecting Cross-sample Cell Developmental Paths (T3).} 
\Ex{1} analyzed paths formed by the selected cell types based on frequency, as indicated by the bar chart width to the left of each row in the \pathcomparison. 
\Path{2} showed the highest frequency \Subfig{\ref{fig4}}{C}{3}, indicating its prevalence and suggesting it might represent a key developmental path in embryonic development. 
Additionally, \Path{2} exhibited optimal alignment in cell spatial distribution across nodes, with the contour projections of its three cell types showing significant overlap on both the $x$- and $y$- axis in the summary visualization \Subfig{\ref{fig4}}{C}{4}.
Moreover, the similarity of spatial distribution in \Path{2} is further supported by the color intensity and height of the links between each pair of nodes \Subfig{\ref{fig4}}{C}{5}. 
The path also presented consistent directionality, as seen from the uniform direction of arcs surrounding each dot in the summary visualization \Subfig{\ref{fig4}}{C}{4}.
Based on these observations, \Ex{1} chose \Path{2} for further investigation.
Upon clicking the left-most button of \Path{2}, a new tab opened in the \pathcomparison, displaying the list of cross-sample trajectories constituting \Path{2} in \trajectorycomparison \Subfig{\ref{fig4}}{C}{2}. 
Moreover, \Ex{1} noticed that the cell types of \Path{3} showed significant similarity to those of \Path{2} by the width and color intensity bar charts on the top of each column \Subfig{\ref{fig4}}{C}{1}, suggesting potentially related developmental processes.

\textbf{Gene Function Analysis (T4, T5).}
When exploring the cross-sample trajectories \Subfig{\ref{fig4}}{C}{2}, \Ex{1} found that \Subpath{2}{13} exhibited the highest alignment in cellular distribution, as indicated by the most consistent contour projections on the summary visualization \Subfig{\ref{fig4}}{C}{2}. 
Therefore, \Subpath{2}{13} was selected for gene function analysis.
The results highlighted significant gene functions in neuronal apoptosis (\eg GO:0000122) and cell proliferation control (\eg GO:0042127), suggesting its potential role in brain specialization \Fig{\ref{fig1}}{D}.
Conversely, \Subpath{2}{8} and \Subpath{2}{15} had few or even no enriched gene functions, indicating \Subpath{2}{8} represents an early phase of brain development while \Subpath{2}{15} resides closer to the developmental starting point.

In this case, \Ex{1} explored cross-sample cell developmental paths associated with brain cells and identified paths that exhibit biological evolutionary relationships through spatiotemporal patterns. 

\subsection{Case Study II: Identifying Cross-sample Cell Developmental Branch Patterns in Mouse Midbrain Dataset}

\Ex{5}, a medical expert who specializes in neurology and frequently uses bioinformatics tools for \scrna data analysis, utilized \system to study mouse neural development in Dataset 2.

\begin{figure}[t]
	\centering
	\includegraphics[width=\linewidth]{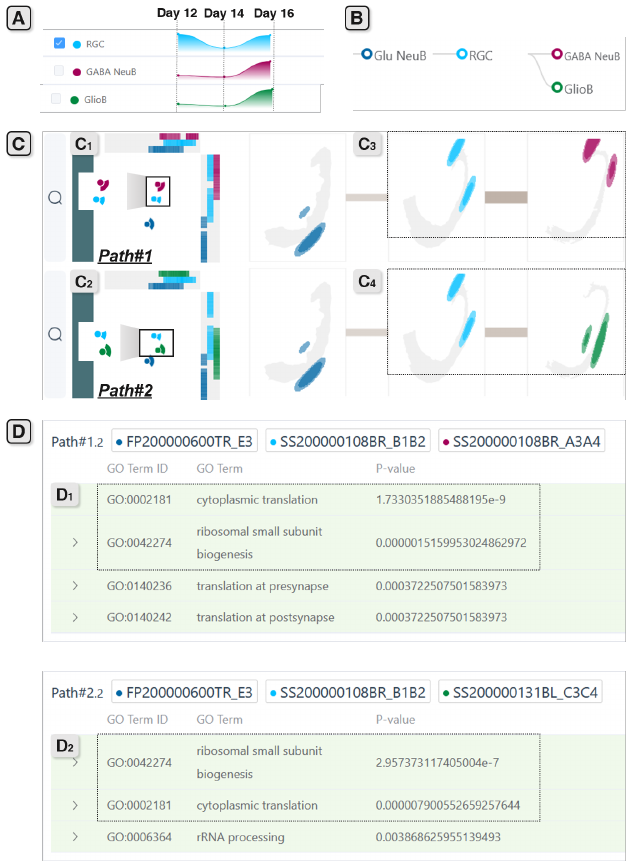}
	\caption{
		Identifying cell developmental branches in Case Study II. 
		(A) The \celltable presents RGC, GABA NeuB, and GlioB across three stages: Day 12 to Day 16. The quantity of RGC began to decrease at D14, while the quantities of GABA NeuB and GlioB increased from D14.
		(B) The \pathselection shows the predicted links related to RGC, indicating a potential branch relationship between these three cell types.
		(C) The \pathcomparison presents cell distribution along \Path{1} and \Path{2}, confirming their spatial proximity.
		(D) The \genefunction reveals similar gene expression profiles for the two paths, confirming their branching evolutionary relationship.
	}
	\label{fig5}
	\vspace{-1em}
\end{figure}

\textbf{Identifying Cell Developmental Branches (T1, T2).}
Firstly, \Ex{5} observed the quantity of RGC decreased at Day 14 while GABA NeuB and GlioB increased from Day 14 to Day 16 \Fig{\ref{fig5}}{A} through corresponding line charts. 
Selecting RGC, \Ex{5} found multiple predicted links from RGC to GABA NeuB and GlioB in \pathselection \Fig{\ref{fig5}}{B}, suggesting RGC may differentiate into GABA NeuB and GlioB simultaneously. 
Moreover, \Ex{5} observed similar contour maps (Fig.\ref{fig5}(C{\small{3}}, C{\small{4}})) and similar direction of arcs surrounding dots corresponding to GABA NeuB and GlioB (Fig.\ref{fig5}(C{\small{1}}, C{\small{2}})), which represents the similar developmental directions among these two cell types and supported RGC differentiation into GABA NeuB and GlioB. 

\textbf{Validating Cell Developmental Branches (T3, T5).}
To test this hypothesis, he conducted gene function analysis in \genefunction on two trajectories of these paths \Subpath{1}{2} and \Subpath{2}{2}. The results revealed that the gene functions expressed in both paths were highly consistent, primarily focusing on cytoplasmic translation (GO:0002181) and ribosomal small subunit biogenesis (GO:0042274) (Fig.\ref{fig5}(D{\small{1}}, D{\small{2}})). These gene functions are closely associated with the rapid proliferation of neural cells, thereby further supporting the hypothesis that RGC differentiates into both GABA NeuB and GlioB.

In summary, \Ex{5} discovered that a cell type RGC can differentiate into two distinct cell types, GABA NeuB, and GlioB, during the midbrain development of mouse embryos. 
He further verified the biological significance of this developmental path in terms of gene function, indicating that differentiation patterns are not limited to a one-to-one relationship but extend to multiple cell types.

\subsection{Gene Function View}

The \genefunction presents the results of gene function analysis in table formats(\textbf{T5}) \Fig{\ref{fig1}}{D}. 
The results include the GO term ID, GO term description, and corresponding p-value, indicating the gene function's significance.
To highlight significant functions, rows with p-values less than 0.05 are marked in green. 
When users expand a row, \genefunction will present a list of all genes associated with the selected GO term, facilitating further analysis based on this gene list.

\section{Evaluation}
In our evaluation process, we first conducted a multi-metric quantitative analysis to assess the performance of our proposed GNN-based predictive model and then validated the practical usefulness and effectiveness of \system through qualitative expert interviews.

\newcolumntype{C}{>{\centering\arraybackslash}X}

\begin{table*}[t]
	\centering
	\begin{threeparttable}
		\caption{
			The evaluation results of our proposed model on five datasets and 5 runs.
		}
		\label{table1}
		\setlength{\tabcolsep}{4pt} %
		\renewcommand{\arraystretch}{1} %
		\begin{tabularx}{0.84\textwidth}{Xccccc}
			\toprule
			Dataset & Our Model & Our Model - GAT & Our Model - Fusion & GraphSAGE \cite{graphsage} & GAT \cite{gat2018petar} \\ 
			\midrule
			Mouse-Embryo \cite{dataset2022chen} & 
			\textbf{88.56 $ \pm $ 1.46} & 
			85.72 $ \pm $ 0.69 & 
			87.83 $ \pm $ 0.46 & 
			\underline{88.42 $ \pm $ 0.18} & 
			87.73 $ \pm $ 0.91 \\ 
			Zebrafish-Embryo \cite{zebrafish2018farrell} & 
			\textbf{76.91 $ \pm $ 0.25} & 
			75.12 $ \pm $ 0.64 & 
			75.81 $ \pm $ 0.42 & 
			74.42 $ \pm $ 0.80 & 
			\underline{76.62 $ \pm $ 0.33} \\
			Schiebinger2019 \cite{Schiebinger2019} & 
			\textbf{71.98 $ \pm $ 1.49} & 
			70.75 $ \pm $ 0.54 & 
			\underline{71.16 $ \pm $ 1.16} & 
			70.26 $ \pm $ 1.59 & 
			70.99 $ \pm $ 1.81\\ 
			CiteSeer \cite{citeseer} &
			\textbf{81.53 $ \pm $ 0.36} & 
			80.59 $ \pm $ 1.12 & 
			\underline{80.87 $ \pm $ 0.54} & 
			79.39 $ \pm $ 1.38 & 
			80.16 $ \pm $ 1.63 \\ 
			CoraFull \cite{corafull} &
			\textbf{88.40 $ \pm $ 0.37} & 
			87.51 $ \pm $ 1.38 & 
			\underline{88.23 $ \pm $ 0.88} & 
			87.50 $ \pm $ 1.45 & 
			87.42 $ \pm $ 1.46 \\ 
			
			\bottomrule
		\end{tabularx}
		
		\begin{tablenotes}
			\small
			\renewcommand{\thempfootnote}{} %
			\setlength{\leftskip}{0pt} %
			\item[] \textbf{Bold} indicates the best performance; \underline{underlined} indicates the second-best.
		\end{tablenotes}
		
	\end{threeparttable}
	\vspace{-1em}
	\mbox{}
\end{table*}

\subsection{Model Evaluation}
We conducted ablation studies to validate the effectiveness of each module in our model and performed comparative experiments to establish its superiority over existing methods.
We also tested the time consumption of our model in completing common tasks to prove its efficiency and hyperparameter settings to prove the rationality of our hyperparameter choices, with details provided in Supplement Materials.

\textbf{Experiment setup.}
In our ablation study, we isolated two key modules of our model: GAT and Fusion. Specifically, we replaced the GAT with a GCN to assess the benefits of attention-based aggregation and replaced the Fusion module into a weighted summation module to examine its impact on feature integration. 
To demonstrate the superiority of our proposed model in our specific application domain, we compared its performance against standard GAT \cite{gat2018petar} and GraphSAGE \cite{graphsage} models.
We utilized the Mouse-Embryo Dataset used in prior analyses and two multi-sample \scrna datasets, Zebrafish-Embryo \cite{zebrafish2018farrell} and Schiebinger2019 \cite{Schiebinger2019}.
Additionally, we utilized two widely adopted GNN benchmarks, CoraFull \cite{corafull} and CiteSeer \cite{citeseer}, to ensure the generalizability of our model.
Details about the datasets is provided in Supplement Materials.
We computed the accuracy (ACC) metric and conducted five repeated experiments to obtain the standard deviation.

\textbf{Evaluation Criteria.}
For the Mouse-Embryo dataset, we collaborated with domain experts to annotate 118 labels for cross-sample cell developmental trajectories, which served as the reference background.
In the case of the Zebrafish-Embryo and Schiebinger2019 datasets, we adopted their inherent differentiation trajectories as ground truth. 
As for the CiteSeer and CoraFull datasets, their original graph structures served as ground truth.
Then, we separately apply our model to edge-less graphs extracted from datasets to predict missing edges. 

\textbf{Results.}
Table \ref{table1} presents the evaluation results of our proposed model on five datasets and averaged over five independent runs, demonstrating its superior performance.
Notably, the model achieves an accuracy  of $88.56 \pm 1.46 $ on the Mouse-Embryo dataset, aligning well with expert expectations for accuracy.
Ablation studies confirm that the full model outperforms its ablated variants, while comparative experiments show it surpasses baseline models.
Additionally, on the CiteSeer and CoraFull datasets, our model exhibits robust performance, highlighting its strong generalization across large-scale datasets. 
In summary, our model's accuracy meets user requirements across diverse datasets while demonstrating robustness and effectiveness.

\subsection{Expert Interview}

To validate the usefulness and effectiveness of \system, we conducted a semi-structured interview with six domain experts over 70 minutes. These experts included the four \Ex{1}-\Ex{6} introduced in Section \ref{sec:section3.2} and two additional experts \Ex{7}, \Ex{8} from different institutions who had experience using tools to construct cell developmental trajectories based on \scrna data.
We first introduced the background of our research (15 minutes), followed by an introduction to the dataset used and the usage step of our system (15 minutes). Then, we asked the experts to complete the exploration process for both Dataset 1 and Dataset 2, with each task requiring at least 20 minutes. Finally, we had a semi-structured interviews with them to collect their feedback (20 minutes). 

\textbf{Usefulness.} 
\Ex{1} to \Ex{4} acknowledged that the data processing and workflow of \system align with the conventions of biology field. 
Our workflow follows standardized biological analysis practices for constructing cell developmental trajectories, analyzing cell developmental paths, and identifying enriched gene functions. 
Our innovation lies in introducing a graph-based prediction model that can capture and preserve the non-linear structure and dynamics of \scrna samples. 
As \Ex{2} noted: \textit{``GNN propagates messages through edges to capture both local cell characteristics and global inter-sample patterns''}.
Moreover, by integrating cellular features, inter-cellular relationships, and cross-sample heterogeneity, our proposed model effectively bridges local details with a global perspective, thereby significantly enhancing predictive accuracy.
Additionally, our system enables biologists to construct developmental trajectories and interpret their biological relationships through multi-dimensional evidence, such as gene expression patterns, cell-cell links, and cellular spatial distribution dynamics.

\textbf{Visualization and Analytical Strengths.} 
All experts praised the visualization design of \system as a key strength.
The \pathselection, which extracts high-frequency paths into a hierarchical tree structure, was highlighted by \Ex{4} as: \textit{``This structure improves the readability of complex trajectory networks by presenting cell relationships with frequency and distance to the core cell. Moreover, it also displays characteristics of entire high-frequency paths, helping me understand developmental relationships and their significance.''}
The contour map abstraction and multi-row layout significantly enhance cross-sample analysis by enabling multi-dimensional comparisons of cell populations in a path. As highlighted by \Ex{5}: \textit{``This approach facilitates the identification of cell populations related to specific tissues while supporting both horizontal and vertical comparisons.''}
Specifically, horizontal comparisons reveal cellular changes across different developmental paths, while vertical comparisons highlight variations in cell distributions at corresponding positions within diverse trajectories. \textit{``This dual-axis analytical framework provides a comprehensive view of cellular dynamics and differentiation patterns''}, noted by \Ex{7}.
The summary visualization for each row, achieved through the projection of cell states onto $x$- and $y$-axis and the summary of distribution maps, enables users to compare cellular patterns across paths intuitively. 
Such an overview-to-detail analytical framework provides a comprehensive understanding of cellular dynamics and differentiation patterns.

\textbf{Suggestion.} 
\Ex{5} suggested incorporating a pseudo-time ordering algorithm \cite{dtflow2021wei} to calculate the direction of cellular development, thereby identifying the developmental direction of specific cell types more clearly. This would facilitate users in assessing the continuity of developmental directions among multiple cells. 
Additionally, \Ex{3} suggested that the system should support the selection of multiple core cell types for comparative analysis, thereby revealing interaction patterns associated with these different core cell types.
\Ex{2} proposed that the system can directly provide differential gene expression analysis results for each path, enabling rapid identification of key differentially expressed genes and supporting hypothesis validation in exploration processes.

\section{Discussion}

In this section, we discuss the design implications, stability of our \pathcomparison, and limitations and future work.

\textbf{Design Implications.} 
Our workflow abstracts complex biological subjects into computable structures for analysis.
First, cell developmental trajectories are presented as graphs, where nodes represent specific cell populations and edges represent cell developmental trajectories, which enables the prediction of cross-sample cell developmental trajectories based on GNN.
Second, spatial distributions of cells from multiple samples are simplified into contour maps highlighting key features such as density peaks, tissue boundaries, and overlapping regions while facilitating cross-sample comparisons.
These abstractions convert biological data into computable formats such as graphs and maps, providing an intuitive analysis workflow and visualization that facilitates users' understanding of these complex datasets.

\textbf{Scalability.}
For analyzing cross-sample cell developmental trajectories, we used the Mouse-Embryo dataset, which spans a few developmental developmental stages. Due to its temporal resolution, predicted paths of the dataset are typically four steps or shorter.
To address more complex \scrna datasets, we collected two large-scale datasets Zebrafish-Embryo and Schiebinger2019 to evaluate the model's scalability. 
Additionally, we validated the model's generalizability using widely used GNN benchmark datasets CiteSeer and CoraFull.
\system currently implements multi-row visualization for analyzing cross-sample cell developmental trajectories. 
Considering longer or more paths, \system allows users to explore by scrolling horizontally or vertically. 
Additionally, our system supports sorting multiple developmental paths by their frequency, enabling users to quickly identify high-frequency paths and improve analysis efficiency.

\textbf{Limitations and Future Work.} 
Currently, the limited publicly accessible multi-sample \scrna datasets restrict our ability to explore various relationships across samples. 
To address this, we can incorporate heterogeneous samples from diverse origins to enhance the robustness of our system, such as those describing the same subject but originating from different datasets (\eg samples from different laboratories or batches of the same disease).
In addition, we identified that the cell contour estimation based on Gaussian Kernel Density Estimation generated density estimates that exceeded the actual distribution boundaries of the samples, especially in sparse regions or at the edges. This limitation disrupts users' manual comparison of contour overlaps across samples, leading to misinterpretations of cell distribution overlaps. 
To mitigate this, we plan to implement overlay scatter plots of raw cell distributions as reference layers to help users distinguish between ``true'' and ``estimated'' contour regions. 

\section{Conclusion}

We introduce \system, a visual analytics system with a GNN-based model to assist biologists in predicting and exploring cross-sample cell developmental trajectories for multi-sample \scrna datasets. 
Our proposed multi-row visualization with contour maps facilitates the validation of biological evolutionary relationships of candidate paths through the overview of cellular spatial distributions.
To assess the performance of our model and the effectiveness of our system, we conducted a quantitative evaluation of our model's performance and performed qualitative validation of our system's effectiveness through two case studies and interviews with eight experts.

\acknowledgments{
This work was supported by National Natural Science Foundation of China (Nos.62172289). 
}

\section*{Supplemental Materials}

All supplemental materials are available on GitHub at \url{https://doi.org/10.17605/OSF.IO/ETFJ2}, including dataset descriptions and extended results.

\bibliographystyle{abbrv-doi-hyperref}

\bibliography{reference}

\end{document}


\maketitle

\section*{Appendix A: Derivation of Eq. (5)}
In the dual analysis framework, taking modifying $S$ to $S'$ as an example, the updating of X is formulated as:
\begin{equation}
    \begin{aligned}
    \label{eq:x_update}
        X' & = &  \underset{X} {\arg\min} & & &  \|p_s(X) - S'\|^2 \\
           & & \mathrm{s.t.}  & & &  p_s = \underset{p_s}{\arg\min}\, O_s(p_s, X).
    \end{aligned}
    \tag{3}
\end{equation}
When Multidimensional Scaling (MDS) is used as the projection method, since MDS directly provides the projection results rather than relying on a projection function, Eq.~(\ref{eq:x_update}) is transformed into:
\[
    \begin{aligned}
    \label{eq:x_update_mds}
        X' &=& \underset{X}{\arg\min} & & & \sum_{i} {(s_i - s'_i)^2}  \\
           & &\mathrm{s.t.} & & & S = \underset{S}{\arg\min}\, \sum_{i,j} \left\| \sqrt{(x_i - x_j)^2} - \sqrt{(s_i - s_j)^2} \right\|^2.
    \end{aligned}
\]
Assuming that ${X}''=\{x_1'',x_2'',..\}$ represents the optimal solution for the optimization problem:
\[
\underset{X}{\arg\min}\, \sum_{i,j} \left\| \sqrt{(x_i - x_j)^2} - \sqrt{(s'_i - s'_j)^2} \right\|^2.
\]
Then, in Eq.~(\ref{eq:x_update}), let $x_i=x_i''$ and $s_i=s_i'$ for all $i$.
This choice of values guarantees that the constraint in Eq.~(\ref{eq:x_update}) is satisfied, while the objective in Eq.~(\ref{eq:x_update}) is minimized (\ie, zero).
It means that the optimization problem above is equal to optimizing Eq.~(\ref{eq:x_update}).
When \( X' = XW \), the optimization problem above is transformed into Eq.~(\ref{eq:sirius}):
\begin{equation}
\begin{aligned}
    \label{eq:sirius}
    W &= \underset{W}{\arg\min} \sum_{i,j} \left\| \sqrt{((XW)_i - (XW)_j)^2} - \sqrt{(s'_i - s'_j)^2} \right\|^2 \\
      &= \underset{W}{\arg\min} \sum_{i,j} \left\| \sqrt{(x_i W - x_j W)^2} - \sqrt{(s'_i - s'_j)^2} \right\|^2.
\end{aligned}
\tag{5}
\end{equation}

\clearpage
\section*{Appendix B: Proof of Theorem 1}
\newtheorem{theorem}{Theorem}
\begin{theorem}
\label{theorem:opt}
Using an invertible neural network as the projection function $p_s(\cdot)$. When $S$ is modified to $S'$, let $X'_{\text{opt}}$ be the optimal updated X obtained by optimizing Eq.~(\ref{eq:x_update}).
Then we have
$$X'_{\text{opt}} = p_s^{-1}(S').$$
\end{theorem}
\begin{proof}
Since $ p_s $ is invertible, there exists a unique $ X = p_s^{-1}(S') $ such that $ p_s(X) = S' $. Then we obtain
$$
\| p_s(p_s^{-1}(S')) - S' \|^2 = \| S' - S' \|^2 = 0.
$$
This is the global minimum of $ \| p_s(X) - S' \|^2 $, which is non-negative and zero only when $ p_s(X) = S' $. 
Therefore, $X'_{\text{opt}} = p_s^{-1}(S')$.
\end{proof}

\section*{Appendix C: Proof of Theorem 2}
\begin{theorem}
\label{theorem:comparison}
Using an invertible neural network as the projection method $p_s(\cdot)$.
When $S$ is modified to $S'$, let $X'_{\text{SIRIUS}}$ be the updated X obtained by SIRIUS, and $X'_{\text{inv}}$ be the updated X obtained by our method.
Then we have 
$$\|p_s(X'_{inv}) - S'\|^2 \leq \|p_s(X'_{SIRIUS}) - S'\|^2.$$
\end{theorem}
\begin{proof}
By Theorem~\ref{theorem:opt}, when using an invertible neural network $p_s(\cdot)$, the optimal updated $X'_{\text{inv}}$ satisfies $X'_{\text{inv}} = p_s^{-1}(S')$, and thus $p_s(X'_{\text{inv}}) = S'$, implying
$$\|p_s(X'_{\text{inv}}) - S'\|^2 = 0.$$
Since $\|p_s(X'_{\text{SIRIUS}}) - S'\|^2 \geq 0$ for any $X'_{\text{SIRIUS}}$ produced by SIRIUS, it follows that $\|p_s(X'_{\text{inv}}) - S'\|^2 \leq \|p_s(X'_{\text{SIRIUS}}) - S'\|^2$.
\end{proof}

\clearpage
\section*{Appendix D: Neural Network Structure of Our Method}
\subsection*{D.1 Network structure}
\begin{figure}
    \centering
    \includegraphics[width=\linewidth]{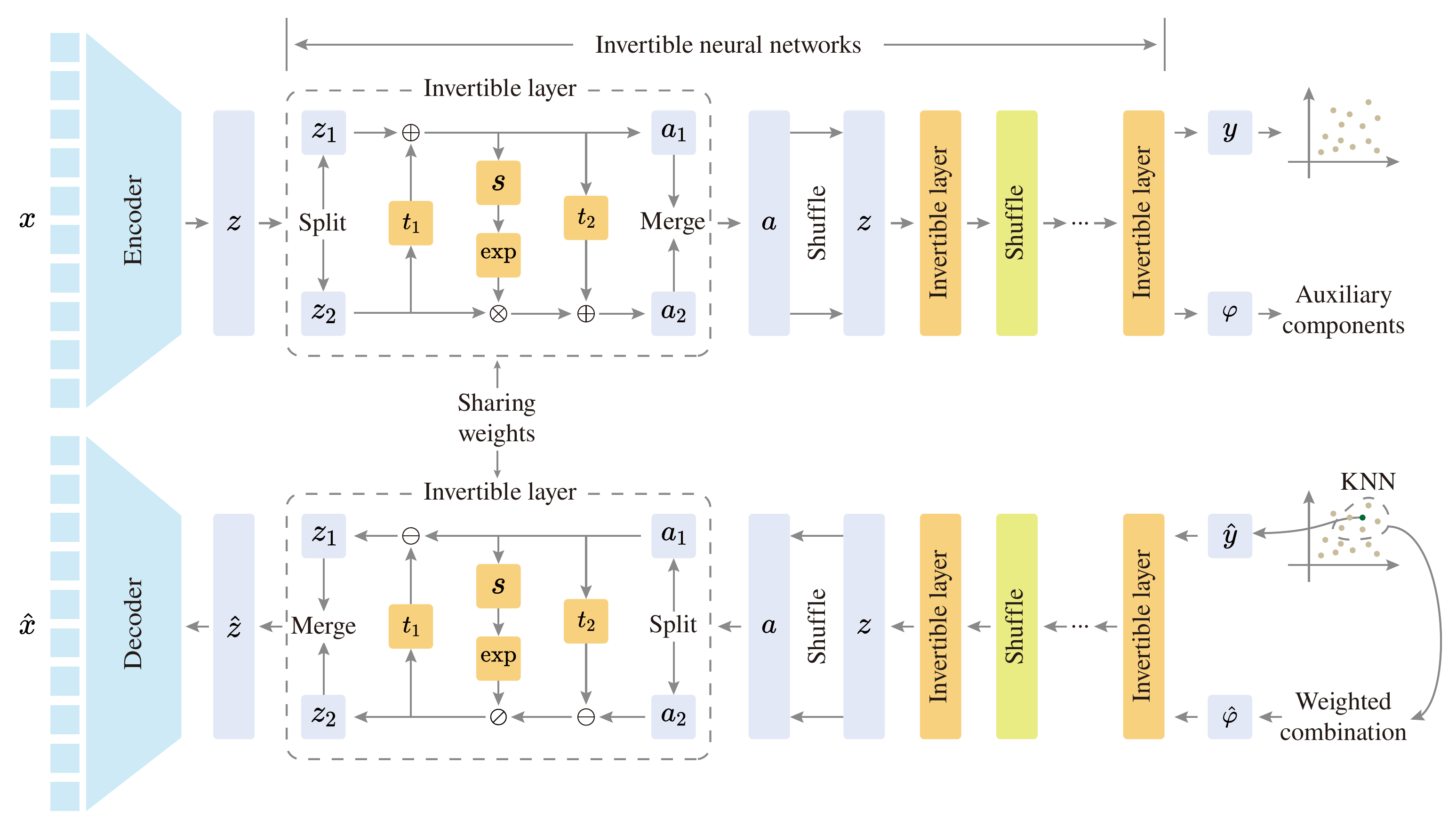}
    \caption{Neural network structure of our method.}
    \label{fig:model}
\end{figure}

\pcheng{Fig.~\ref{fig:model} illustrates our method’s neural network, integrating invertible neural networks (INNs) within an autoencoder. The autoencoder maps input $x$ to latent representation $z$, which INNs transform into a 2D projection $y$ and auxiliary component $\varphi$. Given a new 2D input $\hat{y}$, INNs compute the auxiliary component $\hat{\varphi}$ and invert them to $\hat{z}$, enabling reconstruction of $\hat{x}$.}

\myparagraph{\pcheng{Invertible layer.}}
\pcheng{INNs employ affine coupling layers (ACLs)~\cite{dinh2016density}, where each layer splits input $z$ into $(z_1, z_2)$ and outputs $(a_1, a_2)$. While standard ACLs set $a_1=z_1$, we enhance this through an additive transformation inspired by InvVis~\cite{ye2023invvis}. The forward transformation is expressed as:
\[
\begin{aligned}
    a_1 &= z_1 + t_1(z_2), \\
    a_2 &= z_2 \otimes \exp(s(a_1)) + t_2(a_1).
\end{aligned}
\]
The inverse is:
\[
\begin{aligned}
    z_2 &= (a_2 - t_2(a_1)) \otimes \exp(-s(a_1)), \\
    z_1 &= a_1 - t_1(z_2).
\end{aligned}
\]
Here, $\otimes$ is the Hadamard product, and $s(\cdot)$, $t_1(\cdot)$, $t_2(\cdot)$ are transformation blocks, implemented with fully connected layers.}

\pcheng{To avoid direct alignment between the first two dimensions of input $z$ and the output $y$, we apply shuffle layers inspired by RealNVP~\cite{dinh2016density} with a fixed random permutation $\pi$ between invertible layers.}

\myparagraph{\pcheng{Projection and reconstruction.}}
\pcheng{INNs continuously reduce the dimensionality of $z_1$ while increasing that of $z_2$. In the final layer, $a_1$ becomes the 2D projection $y$, and $a_2$ is $\varphi$. For a new $\hat{y}$, $\hat{\varphi}$ is computed using k-nearest neighbors (KNN) from training data:
\[
\hat{\varphi} = \frac{\sum_{i=1}^{k} \frac{1}{d(\hat{y}, y_{i})} \cdot \varphi_{i}}{\sum_{i=1}^{k} \frac{1}{d(\hat{y}, y_{i})}},
\]
where $d(\cdot)$ is Euclidean distance.}

\subsection*{D.2 Network configurations in the Dual Projection Experiment}

\myparagraph{\pcheng{Architecture details.}}
\pcheng{The autoencoder uses a four-layer fully connected network with batch normalization and ReLU activations for both encoder and decoder. The INNs comprise four invertible layers, with transformation blocks ($s(\cdot)$, $t_1(\cdot)$, $t_2(\cdot)$) as four-layer fully connected networks with batch normalization and ReLU. Hidden dimensions are 64 for MNIST and CIFAR-10, and 256 for the genomic dataset.}

\myparagraph{\pcheng{Training parameters.}}
\pcheng{We train with a batch size of $N/10$ (where $N$ is the dataset size), learning rate of 0.01, and 1000 epochs using the AdamW optimizer and cosine learning rate scheduler. For contrastive learning, we adopt the training protocol of CDR~\cite{xia2022cdr} but set the positive sample sampling range to $k=10$. The reconstruction loss weight $\lambda$ is set to 0.1.}

\clearpage
\section*{Appendix E: More Details of the Dual Projection Experiment}
\myparagraph{\pcheng{Measure definition.}} \pcheng{Trustworthiness ($T$) and Continuity ($C$) evaluate dimensionality reduction quality by assessing preservation of high-dimensional data structure in low-dimensional embeddings. $T$ ensures nearby points in high-dimensional space remain nearby in low-dimensional space. $C$ ensures points close in low-dimensional space are close in high-dimensional space.}

\pcheng{For $n$ data points and neighborhood size $k$:
\begin{equation}
T(k) = 1 - \frac{2}{nk (2n - 3k - 1)} \sum_{i=1}^n \sum_{j \in U_k(i)} (r(i,j) - k),
\end{equation}
where $U_k(i)$ is the set of points among $k$ nearest neighbors of point $i$ in low-dimensional space but not in high-dimensional space and $r(i,j)$ is the rank of point $j$ in point $i$'s high-dimensional neighbors.
}
\pcheng{
\begin{equation}
C(k) = 1 - \frac{2}{nk (2n - 3k - 1)} \sum_{i=1}^n \sum_{j \in V_k(i)} (\hat{r}(i,j) - k),
\end{equation}
where $V_k(i)$ is the set of points among $k$ nearest neighbors of point $i$ in high-dimensional space but not in low-dimensional space and $\hat{r}(i,j)$ is the rank of point $j$ in point $i$'s low-dimensional neighbors.
}

\myparagraph{\pcheng{The influence of k.}} 
\pcheng{In the dual projection experiment, we followed MFM~\cite{ye2025modalchorus} and set k as 30.
We also conducted an experiment to evaluate the influence of k. We tested k=20, 30, and 40 on the MNIST, CIFAR-10, and Genomic datasets.
The results are shown in Tab.~\ref{tab:trustworthiness_continuity}.
As we can see, our method consistently performs better than SIRIUS and is comparable to single projection methods.}

\begin{table}[ht]
\centering
\caption{\pcheng{Trustworthiness (T) and continuity (C) scores of single and dual projection methods for k=20, 30, and 40.}}
\label{tab:trustworthiness_continuity}
\pcheng{
\begin{subtable}{\textwidth}
\centering
\caption{\pcheng{Trustworthiness (T)}}
\begin{tabular}{c|c|ccc|ccc|ccc}
\hline
& \multirow{2}{*}{Methods} & \multicolumn{3}{c|}{MNIST} & \multicolumn{3}{c|}{CIFAR-10} & \multicolumn{3}{c}{Genomic dataset} \\
\cline{3-11}
& & T(20) & T(30) & T(40) & T(20) & T(30) & T(40) & T(20) & T(30) & T(40) \\
\hline
\multirow{2}{*}{Single} & t-SNE & \textbf{0.972} & \textbf{0.963} & \textbf{0.959} & \textbf{0.958} & \textbf{0.951} & \textbf{0.946} & \textbf{0.963} & \textbf{0.957} & \textbf{0.954} \\
& PCA & 0.735 & 0.735 & 0.735 & 0.784 & 0.784 & 0.785 & 0.855 & 0.856 & 0.858 \\
\hline
\multirow{2}{*}{Dual}& SIRIUS & 0.775 & 0.775 & 0.776 & 0.786 & 0.786 & 0.787 & 0.893 & 0.888 & 0.885 \\
& Ours   & \textbf{0.962} & \textbf{0.961} & \textbf{0.960} & \textbf{0.943} & \textbf{0.941} & \textbf{0.940} & \textbf{0.978} & \textbf{0.972} & \textbf{0.966} \\
\hline
\end{tabular}
\end{subtable}
}
\vspace{1em}
\pcheng{
\begin{subtable}{\textwidth}
\centering
\caption{\pcheng{Continuity (C)}}
\begin{tabular}{c|c|ccc|ccc|ccc}
\hline
& \multirow{2}{*}{Methods}  & \multicolumn{3}{c|}{MNIST} & \multicolumn{3}{c|}{CIFAR-10} & \multicolumn{3}{c}{Genomic dataset} \\
\cline{3-11}
& & C(20) & C(30) & C(40) & C(20) & C(30) & C(40) & C(20) & C(30) & C(40) \\
\hline
\multirow{2}{*}{Single} & t-SNE   & \textbf{0.958} & \textbf{0.950} & \textbf{0.945} & \textbf{0.962} & \textbf{0.958} & \textbf{0.955} & \textbf{0.956} & \textbf{0.948} & \textbf{0.943} \\
& PCA & 0.912 & 0.905 & 0.899 & 0.922 & 0.917 & 0.914 & 0.917 & 0.910 & 0.905 \\
\hline
\multirow{2}{*}{Dual}& SIRIUS & 0.892 & 0.888 & 0.886 & 0.897 & 0.895 & 0.893 & 0.922 & 0.919 & 0.918 \\
& Ours   & \textbf{0.951} & \textbf{0.944} & \textbf{0.938} & \textbf{0.953} & \textbf{0.949} & \textbf{0.946} & \textbf{0.959} & \textbf{0.950} & \textbf{0.944} \\
\hline
\end{tabular}
\end{subtable}
}
\end{table}

\clearpage
\section*{Appendix F: Full Qualitative Projection Results}
\begin{figure}
    \centering
    \includegraphics[width=\linewidth]{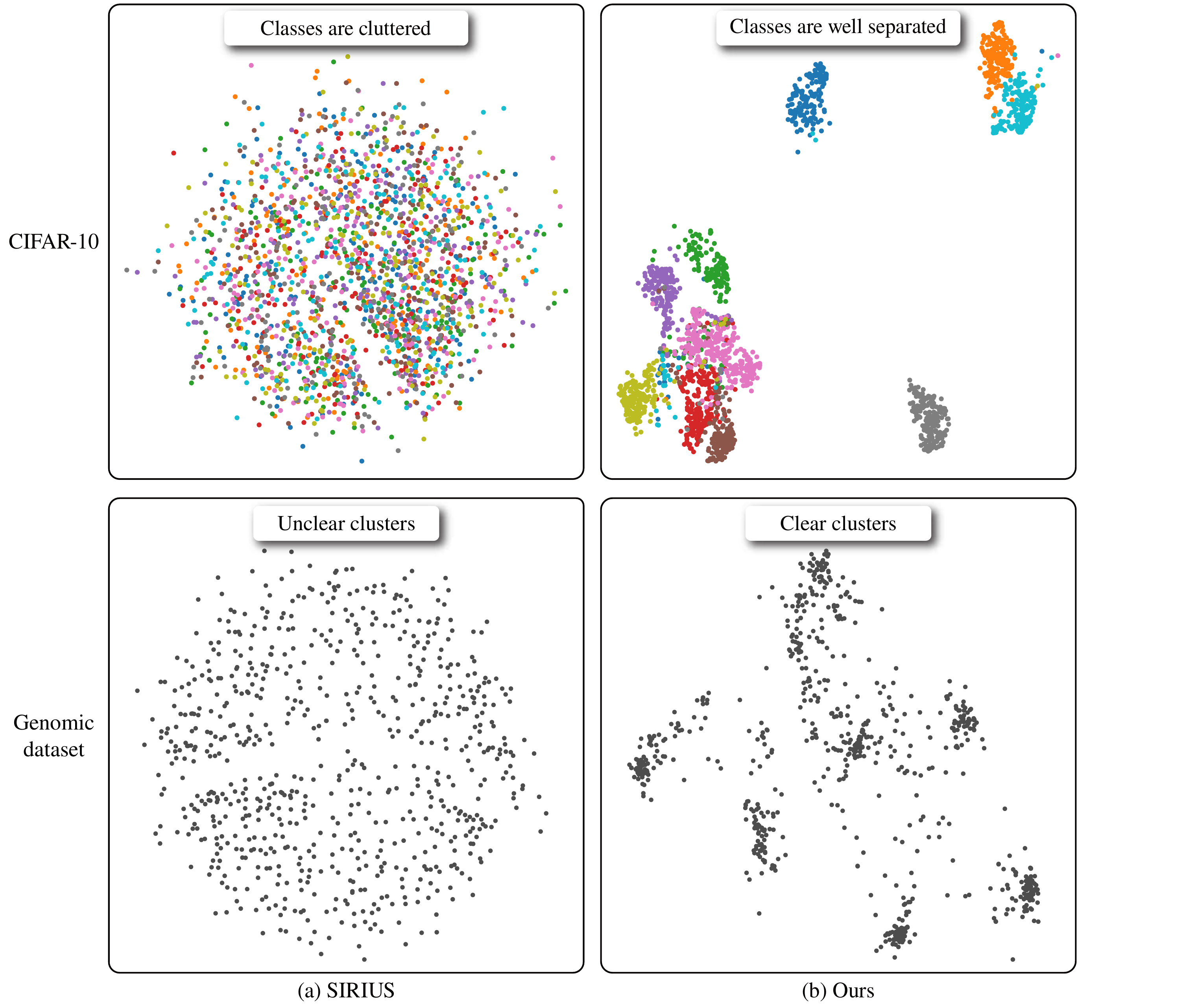}
    \caption{Visual comparison on the CIFAR-10 and genomic dataset.}
    \label{fig:cifar-gene}
\end{figure}

Fig.~\ref{fig:cifar-gene} presents the comparison between SIRIUS and our method on both the CIFAR-10~\cite{krizhevsky2009cifar-10} and genomic datasets, leading to similar conclusions as MNIST. On both datasets, SIRIUS produces densely clustered classes, hindering effective visual exploration. In contrast, our method consistently achieves clear class separation, highlighting its effectiveness in preserving neighborhood relationships.

\clearpage
\bibliographystyle{abbrv-doi-hyperref}

\bibliography{reference}